# Large phase relaxation length in the topological surface states of epitaxial $Bi_{2-\delta}Sn_\delta Te_3$ thin films

Seong Won Cho[1,2], Kwang-Chon Kim[1,3], Seong Keun Kim[1], Byung-ki Cheong[1,2], Jin-Sang Kim[1,2*], and Suyoun Lee[1,2*]

[1]*Electronic Materials Research Center, Korea Institute of Science and Technology, Seoul 136-791, Republic of Korea*

[2]*Department of Nanomaterials, Korea University of Science and Technology, Daejon 305-333, South Korea*

[3]*School of Electrical and Electronic Engineering, Yonsei University, 262 Seongsanno, Seodaemun-Gu, Seoul 120-749, Republic of Korea*

**A topological insulator (TI), a new quantum state featured with the topologically-protected surface state (TSS) originating from its peculiar topology in band structure, has attracted much interest due to academic and practical importance. Nonetheless, a large contribution of the bulk conduction, induced by unintended doping by defects, has hindered the characterization of the unique surface state and the utilization of it into a device. To resolve this problem, we have investigated the transport properties of epitaxial $Bi_{2-\delta}Sn_\delta Te_3$ thin films with varying $\delta$. With the bulk conduction being strongly suppressed, the TSS is separately characterized, resulting in a large phase relaxation length of ~250 nm at 1.8 K, a record-high value in TIs. In addition, the magnetoresistance ratio (MR) has shown a non-monotonic temperature dependence with a maximum value at an elevated temperature depending on $\delta$.**

**These results are associated with the compensation of carriers and, we believe, provide an important step for the application of topological insulators for developing novel functional devices based on the topological surface states.**

A topological insulator (TI) is characterized by conducting surface states with an insulating interior, which are induced by a band inversion due to strong spin-orbit (SO) interaction[1]. The surface states are topologically-protected by time reversal symmetry, which provides a platform for studying the hardly explored subjects so far in the condensed matter physics such as quantum spin Hall (QSH) effect[2] and Majorana fermions[3]. Furthermore, featured by the spin-momentum locked helical Dirac fermions residing on the surface state[4-7], TIs are considered as important ingredients for realizing dissipationless or ultralow-power-consuming devices based on quantum mechanics[8-10]. Due to these properties, archetypal TIs such as $Bi_2Te_3$, $Bi_2Se_3$, and $Sb_2Te_3$ have been intensively studied, with remarkable progresses in the synthesis of high-quality single crystals and thin films and in the characterization of the electrical properties of TIs (for a review, see ref. 1,10-12 and references therein). Nonetheless, the surface conduction is often overwhelmed by the bulk conduction, which originates from unintentional carriers generated by defects such as Bi or Sb vacancies and Te (or Se)/Bi (or Sb) antisites[13,14], making the characterization of the topological surface state (TSS) difficult. To resolve this problem, several approaches have been made to result in a significant suppression of bulk conductivity. Those approaches can be classified into two groups; (1) "defect-suppression scheme", where the formation of defects is prevented by the mixed use of Te and Se due to the increased bonding energy between elements[15-17], (2) "compensation doping scheme" through the inclusion of elements with compensating valence electrons, for example, Sb[18], Ca[5,19], and Sn[14]. Considering that the "defect-suppression scheme" is vulnerable to the phase separation into $Bi_2Se_3$ and $Bi_2Te_3$[20], the "compensation scheme" is regarded as a more viable solution. Indeed, an angle-resolved photoemission spectroscopy (ARPES) study on the effect of doping Sn in $Bi_2Te_3$ demonstrated that Sn is very effective in tuning the Fermi level[14]. In addition, for $Bi_2Te_2Se$, the bulk conduction was shown to be highly suppressed by doping Sn[21]. Motivated by these promising progresses, we move forward herein to explore the magnetotransport properties of Sn-doped TIs. Specifically, we have conducted a systematic study on the magnetotransport properties of epitaxial $Bi_{2-\delta}Sn_{\delta}Te_3$ thin films with varying $\delta$. The bulk conduction has been found to be strongly suppressed with a small concentration of Sn, enabling the characterization of the TSS separately. Moreover, we have also found a few concomitant intriguing features such as the much prolonged phase-relaxation length and the appearance of the maximum MR at an elevated temperature providing an important step toward the application of TIs in functional devices.

$Bi_{2-x}Sn_xTe_3$ films have been grown by Metal-Organic Chemical Vapor Deposition (MOCVD) technique using tri-methylbismuth ($Bi(CH_3)_3$, TMBi), di-isopropyltellurium ($Te(C_3H_7)_2$, DIPTe), and tetramethyltin ($Sn(CH_3)_4$, TMSn) as sources of Bi, Te, and Sn, respectively. Films have been grown on 4$^o$ miscut (001) GaAs substrates and the concentration of Sn have been varied by controlling the flow rate of TMSn at 0, 10, 30, and 60 sccm. The details of growth are described in the Methods section. To quantify the concentration of Sn ($\delta$), the time-of-flight secondary ion mass spectrometry (TOF-SIMS, TOF.SIMS 5, ION-TOF GmbH) has been conducted to show Sn signal systematically changing with the growth conditions[22] but too weak to determine Sn concentration reliably. Auger electron spectroscopy (AES, PHI 700, ULVAC-PHI, INC) has been also tried to detect Sn but in vain, presumably from a low Sn concentration below the detection limit (~1 at. %) of the AES technique[22]. Therefore, as an alternative, we have calculated $\delta$ from the hole concentration obtained by Hall measurement assuming that doped Sn atoms provide holes equal to the sum of free holes and compensating holes required to be combined with free electrons existing in the undoped $Bi_2Te_3$ film. In this way, using the Hall coefficient measured at low magnetic field region (-2~2 T) and at temperature (100 K) high enough to ionize Sn dopants, $x$ has been calculated to be $2.2\times10^{-4}$, $2.8\times10^{-4}$, and $3.7\times10^{-4}$ for flow rate of TMSn at 10, 30, and 60 sccm, respectively.

The crystalline quality of the grown $Bi_{2-x}Sn_xTe_3$ films has been analyzed by high angle annular dark field scanning transmission electron microscopy (HAADF-STEM) and electron back-scattering diffraction (EBSD). Fig. 1(a) shows the HAADF-STEM image of an epitaxial (00*l*) $Bi_{2-\delta}Sn_\delta Te_3$ film, where the $Bi_2Te_3$ (003) lattice fringes are clearly shown. Fig. 1(b) and 1(c) show the highly magnified image of the marked area in Fig. 1(a) and a line scan of intensity, respectively, manifesting the quintuple layers (QLs) composed of alternating Bi and Te layers. Fig. 1(d) shows the fast Fourier transform (FFT) of Fig. 1 (b), clearly indicating the rhombohedral structure (R3m) of the Sn-doped $Bi_2Te_3$ film. Furthermore, Fig. 1(e) shows the in-plane (upper panel) and out-of-plane (lower panel) EBSD images, showing that the grown films are highly-oriented over a large area.

Fig. 2(a) shows the temperature (*T*) dependence of the resistivity ($\rho$) of 100 nm-thick $Bi_{2-\delta}Sn_\delta Te_3$ films. Note that the Sn-doped $Bi_2Te_3$ films show an insulating behavior followed by resistivity saturation below a certain temperature despite a low doping level of Sn, whereas the undoped $Bi_2Te_3$ film shows a metallic behavior before resistivity saturation at low temperature. It clearly indicates that Sn is a very effective dopant for compensating the unintended free electrons existing in the bulk $Bi_2Te_3$ making the surface channel dominant in the conduction process. In Fig. 2(b), the saturated resistivity, represented by $\rho$ at 1.8 K, and the onset temperature ($T_{onset}$) are plotted as a function of $\delta$, where the latter represents the onset temperature of the dominance by the surface channel and is defined as the temperature at which $d\rho/dT$ changes as shown in Fig. 2(a). It is shown that $\rho$(1.8 K) has a maximum around $\delta=2.8\times10^{-4}$ implying the maximal compensation of carriers around the composition, while $T_{onset}$ decreases monotonically with increasing $\delta$.

Fig. 3(a)~(d) show the Hall resistance ($R_{xy}=V_y/I_x$, see Supplementary Information, Fig. S1) as a function of *B* at various temperatures. In contrast to a slight nonlinear behavior and a small *T*-dependent variation for the undoped $Bi_2Te_3$ film, Sn-doped $Bi_2Te_3$ films show quite nonlinear behavior which strongly depends on *T*. At sufficiently high temperatures, all the films show a quite linear behavior but the undoped $Bi_2Te_3$ film is clearly distinguished in that it solely has a negative

slope. This clearly indicates that the nominally undoped-$Bi_2Te_3$ film is actually doped with unintended donors that originate from defects as mentioned above. In contrast, all the Sn-doped $Bi_2Te_3$ films appear p-type, clearly showing the effectiveness of doping Sn for compensating the unintended donors in $Bi_2Te_3$.

Generally, the nonlinear behavior of the $R_{xy}$ vs. $B$ curve implies the existence of multiple types of carriers with distinguished characters. Assuming two types of carriers for simplicity, these curves are fitted to the two-band model described by the following equation (Eq. (1))[23] with four fitting parameters (density ($n$) and mobility ($\mu$) of each type of carrier, $n_1$, $\mu_1$, $n_2$, $\mu_2$) and one constraint ($\sigma = e\sum_i \mu_i n_i$, where $\sigma$ is the electrical conductivity under zero magnetic field).

$$R_{xy} = \frac{R_1 \rho_2^2 + R_2 \rho_1^2 + R_1 R_2 (R_1 + R_2) B^2}{(\rho_1 + \rho_2)^2 + (R_1 + R_2)^2 B^2} B \qquad \ldots (1)$$

, where $R_i=1/n_i e t$ and $\rho_i=1/\mu_i n_i e$ ($t$=thickness of a $Bi_2Te_3$ film, $i$=1, 2). From fitting to Eq. (1) (see Supplementary Information, Fig. S2), it is found that $R_{xy}$ vs. $B$ curves of our Sn-doped $Bi_2Te_3$ films are well described by the two-band model and, as a result, the fitting parameters ($n_1$, $\mu_1$, $n_2$, $\mu_2$) are obtained.

In Fig. 3(e)~(h), sign($n_i$)*log(|$n_i$|) (where $i$=1, 2) are plotted as a function of temperature, where the positive and negative signs mean the p-type and n-type carriers, respectively. Note that $n_2$ changes drastically with Sn concentration in contrast to a little change in $n_1$. Since Sn is expected to supply holes to lower the Fermi level which is located in the bulk-conduction band for the undoped $Bi_2Te_3$ as shown in Fig. 3(a) and 3(e), it seems to indicate that $n_1$ and $n_2$ correspond to the carrier density of the surface channel and bulk channel, respectively. Therefore, this result indicates that the investigated $Bi_{2-\delta}Sn_\delta Te_3$ films are in the underdoped region for $\delta=2.2\times 10^{-4}$, in the near optimally doped region for $\delta=2.8\times 10^{-4}$, and in the overdoped region for $\delta=3.7\times 10^{-4}$, respectively. In addition, the sheet carrier density ($n_{sh}=n_1*t$) of the surface channel is calculated to be about $1\times 10^{13}$ /$cm^2$ for the near optimally doped sample with $\delta=2.8\times 10^{-4}$. And in Fig. 3(i)~(l), the carrier mobilities, $\mu_1$ and $\mu_2$, are plotted as a

function of temperature for each composition. For the undoped $Bi_2Te_3$ film, it is found that the carrier mobility ($\mu_s=\mu_1$) of the surface channel is very high, ~30,000 cm$^2$/Vs, at 1.8 K while that of the bulk channel ($\mu_b=\mu_2$) is about 400 cm$^2$/Vs at low temperatures. Such a high $\mu_s$ makes it possible the observation of the Shubnikov-de Haas (SdH) oscillation as we reported in our previous report[24]. In addition, the observed SdH oscillation was shown to originate from the Dirac electrons by showing the Berry phase of $\pi$. In contrast, for Sn-doped $Bi_2Te_3$ films, $\mu_s$ decreases by about two orders compared to the undoped $Bi_2Te_3$ film while $\mu_b$ increases by about one order at low temperatures. In addition, as a function of temperature, $\mu_b$ shows a peak at a certain temperature depending on the concentration. Looking at Fig. 3(e)~(f), it is found that the temperature at the maximum $\mu_b$ coincides with the temperature where $n_2$ starts to increase drastically. This indicates that the increase in $\mu_b$ with increasing temperature in low temperature region is due to the decrease in the density of charged defects by the compensating holes from Sn dopants. In high temperature region above the temperature at the maximum $\mu_b$ is due to the scattering by phonons. On the other hand, the drastic reduction of $\mu_s$ by doping Sn is not easily explained by the compensation effect. Nevertheless, it is intriguing to note that $\mu_s$ is proportional to the magnitude of $n_2$ regardless of its sign. This observation seems to indicate that the Thomas-Fermi screening[23] plays an important role in the change in $\mu_s$ with Sn concentration in $Bi_{2-x}Sn_xTe_3$.

In Fig. 4(a)~(d), the longitudinal resistance ($R_{xx}$) and magnetoresistance (MR=($R_{xx}(B)$-$R_{xx}(0)$)/$R_{xx}(0)$*100 (%)) of $Bi_{2-x}Sn_xTe_3$ films are plotted as a function of magnetic field ($B$) at various temperatures. For undoped $Bi_2Te_3$ film, MR is found to be as high as 800 %, consistent with the high carrier mobility shown in Fig. 3(i). In contrast, Sn-doped $Bi_2Te_3$ films show much reduced MR of about 150 % regardless of Sn concentration. Furthermore, it is found that Sn-doped $Bi_2Te_3$ films show intriguing temperature dependence of MR, where the magnitude of MR at 9 T depends on $T$ non-monotonically with the maximum value at a certain temperature depending on Sn concentration. This feature will be examined in detail in the later part.

Apart from MR, taking a close look at the behavior in the low *B*-region, a common feature is observed irrespective of Sn concentration; MR curves show a sharp reduction around *B*=0 at low temperatures in contrast to the well-known quadratic dependence on *H* at high temperatures. The behavior at low temperatures is a signature of the weak antilocalization (WAL)[17,25-27]. According to Hikami-Larkin-Nagaoka (HLN) model for the WAL for two-dimensional (2D) conductors[28], the magnetoconductance ($\Delta G_{sh}(B) = G_{xx}(B) - G_{xx}(0)$) is described by the following equation (Eq. (2)).

$$\Delta G_{xx}(B) = \alpha \frac{e^2}{\pi h}[\ln(\frac{h}{8\pi e l_\phi^2 B}) - \psi(\frac{1}{2} + \frac{h}{8\pi e l_\phi^2 B})] \quad \ldots (2)$$

, where $\alpha$, $l_\phi$, $h$, and $\psi(x)$ are a constant representing the number of 2D channels ($\alpha$=1/2 for each channel), the phase-relaxation length, the Planck constant, and the digamma function, respectively. In Fig. 4(e)~(h), $\Delta G_{xx}(B)$ is plotted as a function of *B* at various temperatures and fitting curves by Eq. (2) are also plotted for the data at low temperatures. From the fitting, the values of $\alpha$ and $l_\phi$ can be calculated and are plotted as a function of *T* in Fig. 4(i)~(l). Some intriguing features are found in those figures. First, at low temperatures, the value of $\alpha$ is around unity for Sn-doped $Bi_2Te_3$ films irrespective of Sn concentration whereas it is a few thousands for undoped $Bi_2Te_3$ film. It apparently means that there are two 2D channels dominating the carrier transport for Sn-doped $Bi_2Te_3$ films, consistent with the results of *T*-dependence of $\rho$ shown in Fig. 2(a). Furthermore, it is shown that $\alpha$ increases with increasing *T* above a certain temperature, which is nearly same as $T_{onset}$ in Fig. 2(b) for each composition, bolstering the above explanation. Second, it is shown that $l_\phi$ of Sn-doped $Bi_2Te_3$ films is highly enhanced resulting in about three-fold increase compared to the undoped $Bi_2Te_3$. Since $l_\phi$ is a measure of phase-coherence length of electronic wavefunction, longer $l_\phi$ enables the observation of quantum phase-related phenomena such as Aharonov-Bohm oscillation[29] and the fabrication of functional devices using such quantum phenomena in a large scale. To the best of our knowledge, the obtained value of $l_\phi$~250 nm at 1.8 K for $Bi_{2-\delta}Sn_\delta Te_3$ film with $\delta=2.8\times10^{-4}$ is the record-high value for topological insulators.

In Fig. 5, the curves of MR at 9 T versus temperature for Sn-doped $Bi_2Te_3$ films are plotted. As mentioned above, they show non-monotonic temperature dependence with a maximum at a certain temperature, which decreases with increasing Sn concentration. This feature is peculiar in that MR is mostly observed to increase with lowering temperature, which is associated with the increase of the carrier mobility at low temperature[30]. Furthermore, from the perspective of application, this feature is intriguing in that it can shed a light on the design of a material with high MR at the elevated temperature, which is an important ingredient for developing functional devices based on MR. As to the origin of the intriguing temperature dependence of MR, we believe that it is due to the compensation of carriers. In a simple compensated semiconductor described by the coexistence of two opposite types of carriers (p-type and n-type carriers), $\rho$ is expressed by the following equation (Eq. (3)) in the two-band model[23].

$$\rho = \frac{\rho_1\rho_2(\rho_1+\rho_2)+(\rho_1 R_2^2+\rho_2 R_1^2)B^2}{(\rho_1+\rho_2)^2+(R_1+R_2)^2 B^2} \quad \ldots (3)$$

If the carrier density of electrons and holes are similar, in other words, $R_1 \sim -R_2$, the denominator goes to the minimum, resulting in the strongest dependence of $\rho$ on $B$. An example is $WTe_2$ known as a semimetal with $p/n$ nearly equal to unity, whose MR was reported to be extremely high ($\sim 1.3 \times 10^7$ % at 0.53 K and 60 T) due to the balance of the carrier density of electrons and holes[31]. Note that the maximum MR is shown around the temperature where the electron density at the surface channel is well balanced by the hole density in the bulk as shown in Fig. 3(f)~(h), irrespective of Sn concentration. This clearly indicates that the non-monotonic temperature dependence is closely related with the carrier compensation. This result implies a possibility that the high MR can be achieved at elevated temperatures near room temperature by using dopants with the sufficiently high ionization energy.

To summarize, we have investigated the effect of doping Sn on the transport properties of a topological insulator $Bi_2Te_3$ film grown by MOCVD. From the temperature dependence of resistivity of Sn-doped $Bi_2Te_3$ films, it is found that Sn is an effective dopant for suppressing the unwanted bulk conduction with a small amount of Sn concentration. For Sn-doped $Bi_2Te_3$ films, the magnetic field dependence of the Hall resistance is shown to be highly nonlinear meaning the coexistence of at least two types of carriers, from which we have characterized the carriers in the topological surface states as possessing the carrier density of $\sim 1\times 10^{13}$ /cm$^2$ (at 1.8 K) and the carrier mobility of 300~400 cm$^2$/Vs (at 1.8 K) weakly depending on Sn concentration. In addition, the longitudinal resistance of Sn-doped $Bi_2Te_3$ films at low temperatures is found to be well described by the weak antilocalization. From this, it is found that the phase relaxation length of $Bi_{2-\delta}Sn_\delta Te_3$ film with $\delta=2.8\times 10^{-4}$ has a record-high value of about 250 nm at 1.8 K, which is about three times of that of the undoped $Bi_2Te_3$ film. And, finally, we have found that MR of Sn-doped $Bi_2Te_3$ films show non-monotonic temperature dependence with the maximum at a certain temperature depending on Sn concentration, at which temperature the carrier density of electrons and holes are well balanced. We believe that these results not only advance the characterization of the topological surface states by suppressing the bulk conduction, but also shed a light on the application of a topological insulator to the development of novel devices based on the phase coherence and high MR at elevated temperatures.

**Methods**

Sn-doped $Bi_2Te_3$ films have been grown using Metal-Organic Chemical Vapor Deposition (MOCVD). We have used (001) GaAs substrates with miscut angle of 4$^o$ along [110] direction. Immediately prior to loading the substrate into the reactor, the GaAs substrates were cleaned with HCl (18 %) for three minutes followed by rinsing in deionized water. The growth temperature is 360 $^o$C and high purity $H_2$ gas has been used as the reactant and carrier gas. Tri-methylbismuth (Bi(CH$_3$)$_3$, TMBi), di-isopropyltellurium (Te(C$_3$H$_7$)$_2$, DIPTe) and tetramethyltin (Sn(CH$_3$)$_4$,TMSn) have been used as the sources of Bi, Te and Sn, respectively. The canisters containing source materials have been kept in cold reservoirs. More information about the growth of epitaxial $Bi_{2-\delta}Sn_\delta Te_3$ thin films can be found elsewhere[22]. For the characterization of electrical properties, the films were patterned to the Hall-bar geometry to have various length and width by photolithography followed by wet etching in the solution of $HNO_3$:$HCl$:$H_2O$=23.1:6.9:70. A photograph of a device is presented in the Supplementary Information (Fig. S1).

The structural properties and the microstructure of the films have been examined by x-ray diffraction (XRD; D8 ADVANCE, BRUKER) and by high-angle annular dark field scanning transmission electron microscopy (HAADF-STEM; TITAN, FEI). The crystal orientation of the films has been analyzed by electron back-scattering diffraction (EBSD; Hitachi S4300) with Bruker e-Flash software. The measurements of magnetotransport properties have been performed in a commercial cryogen-free cryostat (CMag Vari. 9, Cryomagnetics Inc.) using the standard lock-in technique. The ac current has been supplied by Keithley 6221 current source with the frequency of 17 Hz and the voltage drop has been measured by the SR830 lock-in amplifier (Stanford Research Systems, Inc.). The electrical contacts to the device have been formed by indium.

**Acknowledgments**

This research was supported by the Korea Institute of Science and Technology research program (Grant No. 2E26420). We acknowledge Dr. Hye Jung Chang for HAADF-STEM analysis and Dr. Dong-Ik Kim for EBSD measurement.

**Author contributions**

J. K. and S. L. contributed to the concept design, transport measurements, data analysis, and interpretation. K. –C. K. and S. K. K. contributed to growth of epitaxial Sn-doped $Bi_2Te_3$ thin films. S. W. C. performed the measurement of transport properties of Sn-doped $Bi_2Te_3$ thin films. B. –k. C. contributed to the analysis and writing of the manuscript. All authors participated in discussion and writing the manuscript.

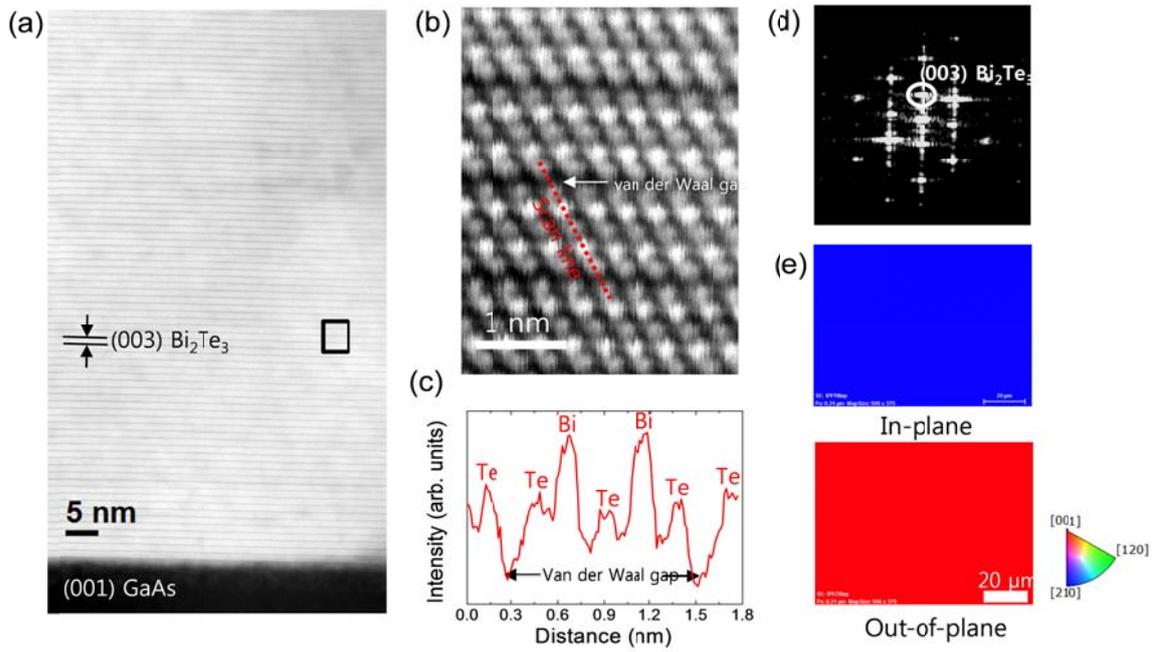

Figure 1. **Structural characterization of Sn-doped $Bi_2Te_3$ films.** (a) High angle annular dark field scanning transmission electron (HAADF-STEM) image of a (00$l$) epitaxial Sn-doped $Bi_2Te_3$ film grown on (001) GaAs substrate (4°-miscut). (b) A magnified image of the square region in (a). (c) The intensity distribution along the red-dotted line in (b), clearly showing Bi and Te atoms consisting of a quintuple layer of $Bi_2Te_3$. (d) Fast Fourier transform (FFT) of (b), which indicates the rhombohedral structure (R3m) of the grown $Bi_2Te_3$ film. (e) In-plane (upper panel) and out-of-plane (lower panel) electron back-scattering diffraction (EBSD) image of a Sn-doped $Bi_2Te_3$ film.

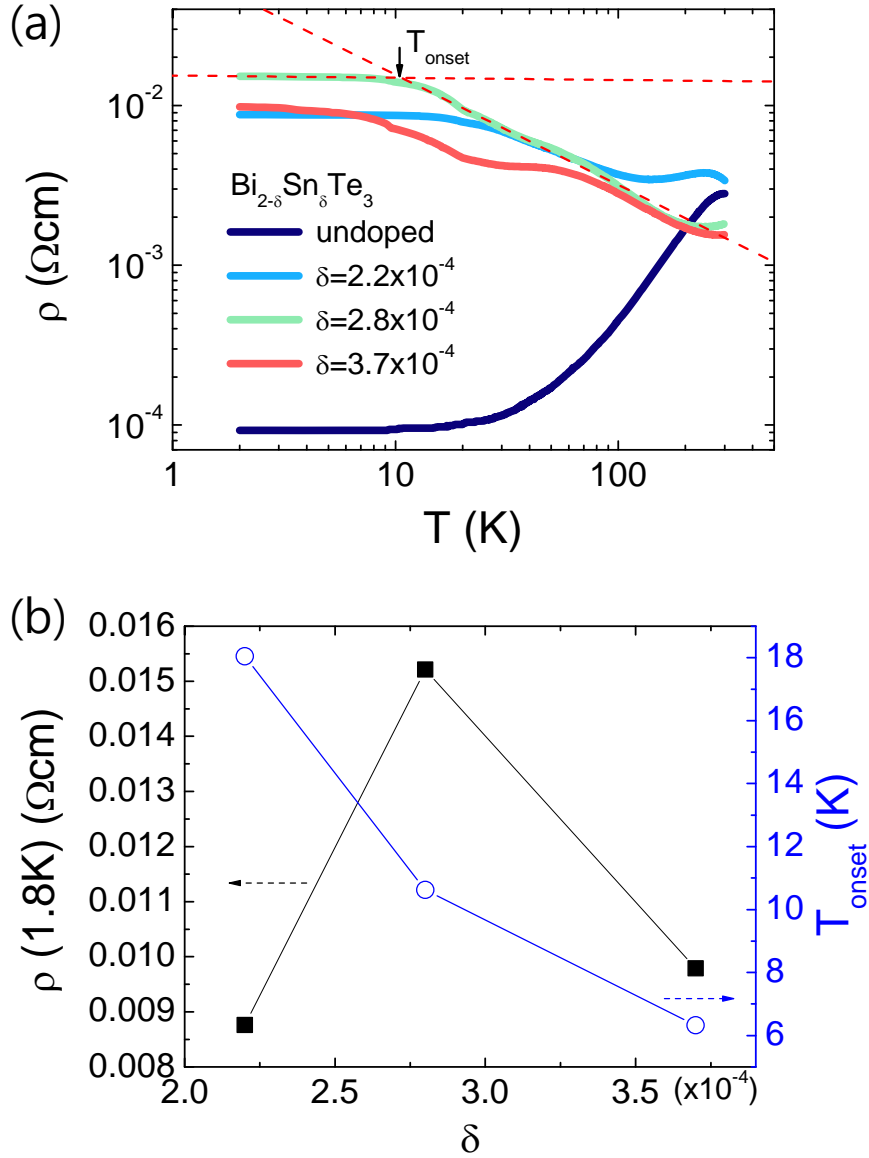

Figure 2. **Temperature (*T*) dependence of resistivity ($\rho$) of 100 nm-thick Sn-doped $Bi_2Te_3$ films**. (a) $\rho$ vs. *T* curves for $Bi_{2-\delta}Sn_\delta Te_3$ films with $\delta=0$ (navy), $2.2\times10^{-4}$ (light blue), $2.2\times10^{-4}$ (light green), and $3.7\times10^{-4}$ (red). $T_{onset}$ is defined graphically for $\delta=2.8\times10^{-4}$ sample. (b) $\rho$(1.8 K) (black) and $T_{onset}$ (blue) are plotted as a function of $\delta$.

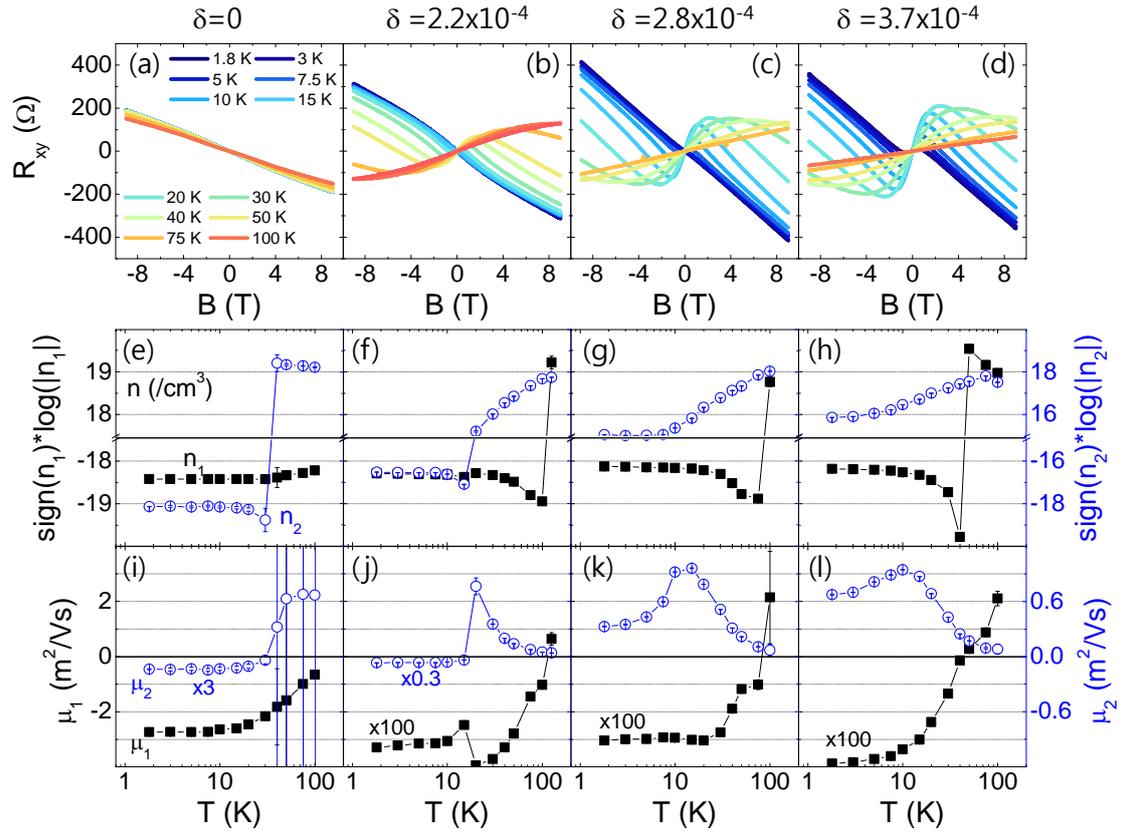

Figure 3. **Sn concentration- and temperature-dependence of Hall resistance ($R_{xy}$) and two-band model analysis of 100-nm thick $Bi_{2-\delta}Sn_\delta Te_3$ films with $\delta=0$, $2.2\times10^{-4}$ (under-doped), $2.8\times10^{-4}$ (optimally-doped), and $3.7\times10^{-4}$ (over-doped), respectively.** (a)~(d) $R_{xy}$ as a function of the applied magnetic field at various temperatures. (e)~(f) Temperature dependence of carrier density ($n_1$ (black, left axis) and $n_2$ (blue, right axis)) of two types of carriers, which are obtained by fitting the $R_{xy}$ data to the two-band model (see the text). Sign($n_i$)*log($|n_i|$) (where $i=1, 2$ and the unit of $n$ is cm$^{-3}$.) are plotted to magnify the change with temperature and Sn concentration. Positive and negative signs represent p-type and n-type carriers, respectively. (i)~(l) Temperature dependence of carrier mobility of two types of carriers. In (e)~(l), error bars are also displayed, which are mostly smaller than the size of the symbol, indicating the goodness of the fitting by the two-band model.

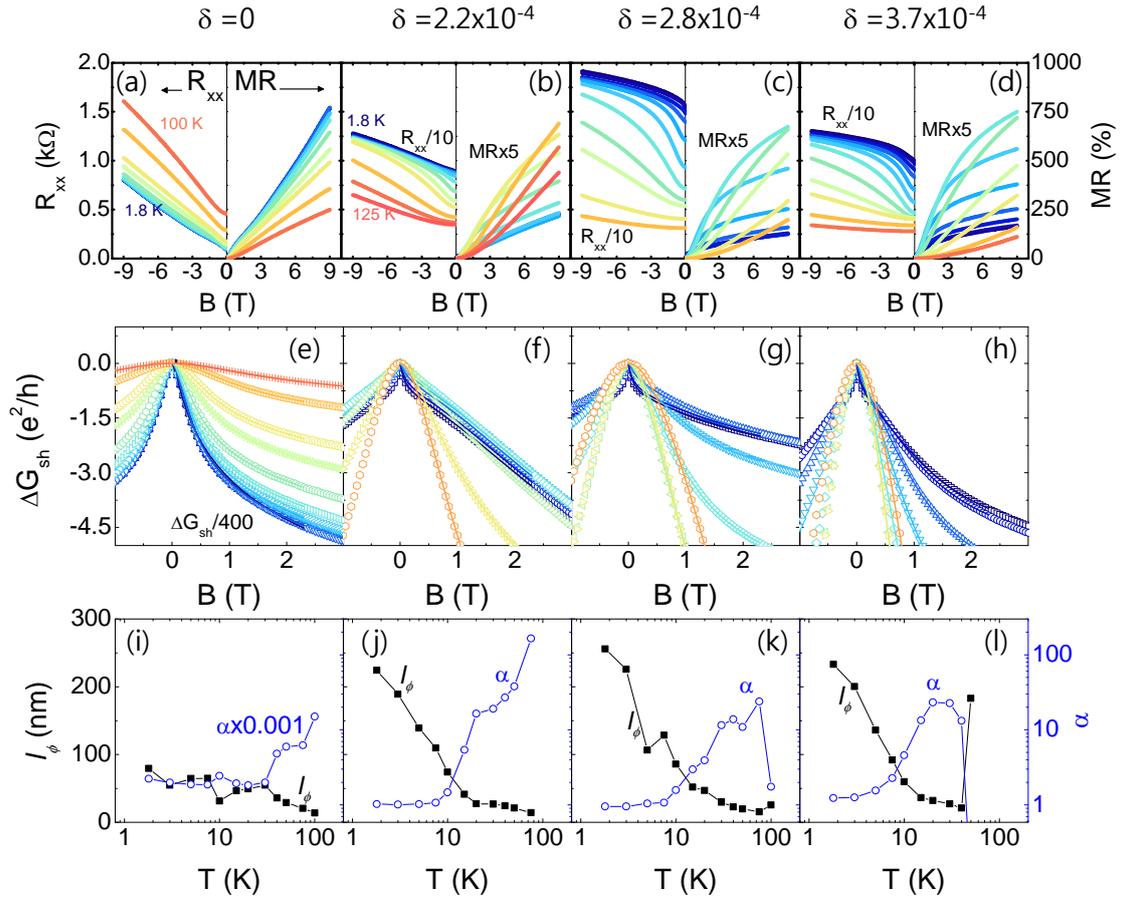

Figure 4. **Composition- and temperature-dependence of longitudinal resistance ($R_{xx}$) and weak-antilocalization (WAL) analysis of 100-nm thick $Bi_{2-\delta}Sn_\delta Te_3$ films with $\delta=0$, $2.2\times10^{-4}$, $2.8\times10^{-4}$, and $3.7\times10^{-4}$, respectively.** (a)~(d) $R_{xx}$ (left panels) and MR (right panels) at various temperatures (1.8 K, 3 K, 5 K, 7.5 K, 10 K, 15 K, 20 K, 30 K, 40 K, 50 K, 75 K, 100 K, and 125 K). The magnetic field is applied perpendicular to the film plane. (e)~(h) $B$-dependence of magnetoconductance ($\Delta G_{sh}=G_{xx}(B)-G_{xx}(0)$, $G_{xx}=1/R_{xx}$) at various temperatures in the low $B$-region (-1~3 T). Symbols and solid lines represent the measured data and the fitting curve by HLN model (see the text), respectively. (i)~(l) Temperature dependence of $\alpha$ and $l_\phi$, which are obtained by fitting to the HLN model.

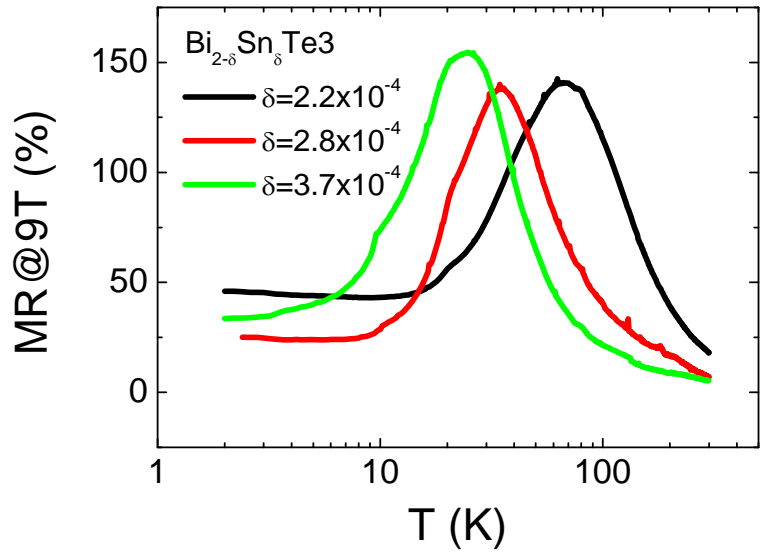

Figure 5. **MR(@9T) of Sn-doped Bi$_2$Te$_3$ films as a function of temperature.**

# [Supplementary Information]

# Large phase relaxation length in the topological surface states of epitaxial Bi$_{2-\delta}$Sn$_{\delta}$Te$_3$ thin films

Seong Won Cho[1,2], Kwang-Chon Kim[1,3], Seong Keun Kim[1], Byung-ki Cheong[1,2], Jin-Sang Kim[1,2*], and Suyoun Lee[1,2*]

[1]*Electronic Materials Research Center, Korea Institute of Science and Technology, Seoul 136-791, Republic of Korea*

[2]*Department of Nanomaterials, Korea University of Science and Technology, Daejon 305-333, South Korea*

[3]*School of Electrical and Electronic Engineering, Yonsei University, 262 Seongsanno, Seodaemun-Gu, Seoul 120-749, Republic of Korea*

**Section 1. A photograph of a device and measurement geometry**

**Section 2. Two-band model fitting of Hall resistance ($R_{xy}$) vs. $B$ curves**

**Section 1. A photograph of a device and measurement geometry**

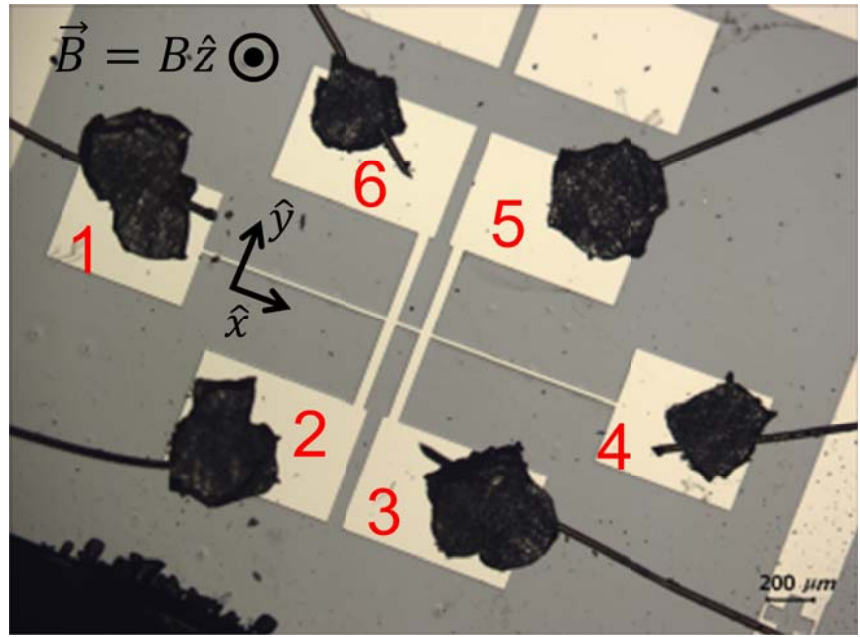

Figure S1. A photograph of a Sn-doped $Bi_2Te_3$ Hall-bar device and the measurement geometry.

## Section 2. Two-band model fitting of $R_{xy}$ vs. $B$ curves

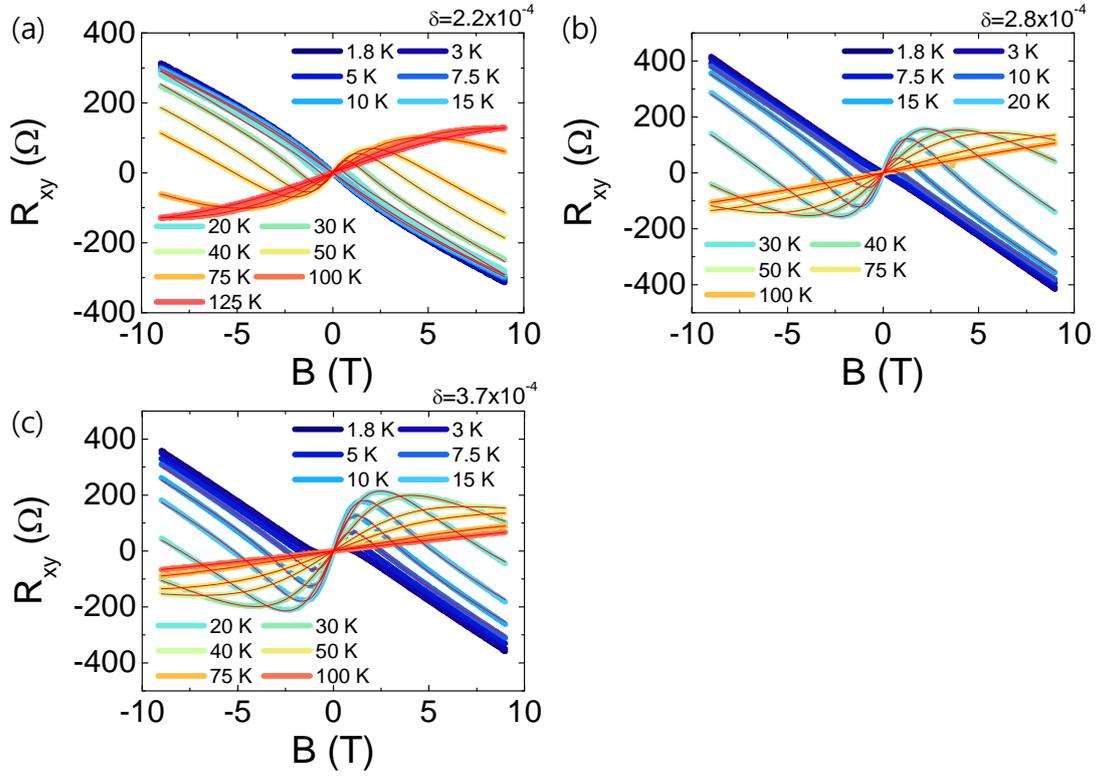

Figure S2. Hall resistance ($R_{xy}$) vs. $B$ curves of 100-nm thick $Bi_{2-\delta}Sn_\delta Te_3$ films with $\delta=2.2\times10^{-4}$ (a), $2.8\times10^{-4}$ (b), and $3.7\times10^{-4}$ (c), respectively, at various temperatures. Thin red lines are fitting curves calculated by the two-band model.